\documentclass[aip, reprint, a4paper, footinbib]{revtex4-1}

\usepackage{graphicx}
\usepackage{color}
\usepackage{amsfonts}
\usepackage{amsmath}
\usepackage{hyperref}
\usepackage{bm}

\begin{document}
\title{Gilbert damping of magnetostatic modes in a yttrium iron garnet sphere}

\author{S.~Klingler}
\email{stefan.klingler@wmi.badw.de}
\author{H.~Maier-Flaig}
\affiliation{Walther-Mei{\ss}ner-Institut, Bayerische Akademie der Wissenschaften, 85748 Garching, Germany}
\affiliation{Physik-Department, Technische Universit\"{a}t M\"{u}nchen, 85748 Garching, Germany}

\author{C.~Dubs}
\author{O.~Surzhenko}
\affiliation{INNOVENT e.V. Technologieentwicklung, 07745 Jena, Germany}

\author{R.~Gross}
\affiliation{Walther-Mei{\ss}ner-Institut, Bayerische Akademie der Wissenschaften, 85748 Garching, Germany}
\affiliation{Physik-Department, Technische Universit\"{a}t M\"{u}nchen, 85748 Garching, Germany}
\affiliation{Nanosystems Initiative Munich, 80799 Munich, Germany}

\author{H.~Huebl}
\affiliation{Walther-Mei{\ss}ner-Institut, Bayerische Akademie der Wissenschaften, 85748 Garching, Germany}
\affiliation{Physik-Department, Technische Universit\"{a}t M\"{u}nchen, 85748 Garching, Germany}
\affiliation{Nanosystems Initiative Munich, 80799 Munich, Germany}

\author{S.T.B.~Goennenwein}
\affiliation{Walther-Mei{\ss}ner-Institut, Bayerische Akademie der Wissenschaften, 85748 Garching, Germany}
\affiliation{Institut f\"{u}r Festk\"{o}rperphysik, Technische Universit\"{a}t Dresden, 01062 Dresden, Germany}
\affiliation{Center for Transport and Devices of Emergent Materials, Technische Universit\"{a}t Dresden, 01062 Dresden, Germany}

\author{M.~Weiler}
\affiliation{Walther-Mei{\ss}ner-Institut, Bayerische Akademie der Wissenschaften, 85748 Garching, Germany}
\affiliation{Physik-Department, Technische Universit\"{a}t M\"{u}nchen, 85748 Garching, Germany}
\date{\today}

	\begin{abstract}
		The magnetostatic mode (MSM) spectrum of a 300\,$\mu$m diameter single crystalline sphere of yttrium iron garnet is investigated using broadband ferromagnetic resonance (FMR). The individual MSMs are identified via their characteristic dispersion relations and the corresponding mode number tuples $(nmr)$ are assigned. Taking FMR data over a broad frequency and magnetic field range allows to analyze both the Gilbert damping parameter~$\alpha$ and the inhomogeneous line broadening contribution to the total linewidth of the MSMs separately.
		The linewidth analysis shows that all MSMs share the same Gilbert damping parameter $\alpha=2.7(5) \times 10^{-5}$ irrespective of their mode index. In contrast, the inhomogeneous line broadening shows a pronounced mode dependence. This observation is modeled in terms of two-magnon scattering processes of the MSMs into the spin-wave manifold, mediated by surface and volume defects.
	\end{abstract}

\maketitle

The ferrimagnetic insulator yttrium iron garnet (YIG) has numerous applications in technology and fundamental research due to its low intrinsic Gilbert damping and large spin-wave propagation length.\cite{Serga2010} It is used as prototypical material in various experiments in spin electronics\cite{Chumak2014, Klingler2015, Ganzhorn2016}
and spin caloritronics\cite{Uchida2008, Xiao2010} and is indispensable for microwave technology.

Recently, YIG spheres attracted attention in the field of quantum information technology.\cite{Huebl2013,Soykal2010, Soykal2010a, Imamoglu2009, Wesenberg2009, Tabuchi2015,Maier-Flaig2016,Klingler2016, Hisatomi2016} For example, strong coupling between magnons and photons in YIG/cavity hybrid systems can be employed for the up- and down-conversion of quantum signals between microwave and optical frequencies, enabling a long-range transmission of quantum information between microwave quantum circuits.\cite{Klingler2016, Hisatomi2016, Zhang2016} Here, the damping of the magnetic excitation plays a crucial role, since it limits the time-scale in which energy and information is exchanged and stored in the magnon-photon hybrid system.

One type of magnetic excitations in YIG spheres\cite{Walker1957,Fletcher1959a, Roschmann1977} are magnetostatic modes (MSMs) which resemble standing spin-wave patterns within the sphere. Although the linewidth of MSMs in YIG spheres has been studied at fixed frequencies in the past,\cite{Roschmann1975, Nemarich1964, Lam1965} the respective contributions of intrinsic Gilbert damping and inhomogeneous line broadening\footnote{Here, inhomogeneous refers to all non-Gilbert like damping mechanisms, while the lineshape remains Lorentzian.} to the total linewidth have not yet been investigated. In particular, it is not evident from the literature, whether different MSMs feature the same or different Gilbert damping.\cite{Holzer1992,Stancil2009}

Here, we report on the study of dynamic properties of multiple MSMs for a 300\,$\mu$m diameter YIG sphere using broadband ferromagnetic resonance. The frequency and magnetic field resolved FMR data allows to separate Gilbert damping and inhomogeneous line broadening of the MSMs. One and the same Gilbert damping parameter $\alpha=2.7(5)\times 10^{-5}$ is found for all MSMs, independent of their particular mode index. However, the inhomogeneous line broadening markedly differs between the observed MSMs. This finding is attributed to two-magnon scattering processes of the MSMs into the spin-wave manifold, mediated by surface and volume defects.

The MSM profiles and eigenfrequencies of a magnetic sphere can be calculated in the magnetostatic approximation $\nabla \times \bm{H}=0$,\cite{Walker1957,Fletcher1959a,Roschmann1977} using the Landau-Lifshitz-Gilbert equation (LLG).\cite{Landau1935, Gilbert2004} The resonance frequencies $\Omega$ of the MSMs are obtained by solving the characteristic equation:\cite{Walker1957,Fletcher1959a,Roschmann1977}
\begin{equation}
\label{char_eq}
	n+1+\xi_0 \frac{\text{d}P_n^m (\xi_0)/\text{d}\xi_0}{P_n^m (\xi_0)}\pm m \nu=0,
\end{equation}
where $\xi^2_0=1+1/\kappa$, $\kappa={\Omega_H}/{\left(\Omega^2_H-\Omega^2\right)}$, $\nu={\Omega}/{\left(\Omega^2_H-\Omega^2\right)}$, $\Omega_H={\mu_0 H_{\text{i}}}/{\mu_0 M_\text{s}}$ and $\Omega= \omega / \gamma \mu_0 M_\text{s}$.
Here, $\gamma=g_J \mu_\text{B}/\hbar$ is the gyromagnetic ratio, $g_J$ is the Land\'{e} $g$-factor, $\mu_\text{B}$ is the Bohr magneton, $\hbar$ is the reduced Planck constant, $\mu_0$ is the vacuum permeability and $M_\text{s}$ is the saturation magnetization. The angular frequency of the applied microwave field is denoted as $\omega = 2 \pi f$. The internal field is given by ${H}_\text{i}= {H}_\text{0} + {H}_\text{ani}+ H_\text{demag}$, where ${H}_\text{0}$ is the applied static magnet field, ${H}_\text{ani}$ is the anisotropy field, and $H_\text{demag}=-M_\text{s}/3$ is the demagnetization field of a sphere. 

The mode profiles of the MSMs have the form of associated Legendre polynomials $P_n^m$, where the localization of the MSMs at the surface is related to the mode index $n \in \mathbb{N}$.\cite{Nemarich1964} The index $|m| \leq n$ corresponds to an angular-momentum quantum number of the MSM,\cite{Wiese1991} where the bar above the mode index $\overline{m}$ is used for indices $m<0$. The index $r\geq0$ enumerates the solutions of the characteristic equation (\ref{char_eq}) for given $n$ and $m$ for increasing frequencies.\cite{Fletcher1959,Fletcher1959a} In total, each MSM is uniquely identified by the index tuple $(nmr)$. For more information and plots of the MSM mode patterns, the review of Ref.~\onlinecite{Roschmann1977} is recommended.

The Gilbert damping parameter phenomenologically accounts for the viscous (linearly frequency-dependent) relaxation of magnetic excitations. Assuming a dominant Gilbert-type damping for all MSM modes, the full linewidth at half maximum (FWHM) $\Delta f^{(nmr)}$ of a MSM resonance line at frequency $f_{\rm res}^{(nmr)}$ is given by:\cite{Kalarickal2006}
\begin{equation}
\label{gilbert_func}
\Delta f^{(nmr)} = 2 \alpha f_\text{res}^{(nmr)} + \Delta f_0^{(nmr)}.
\end{equation}
Here, $\Delta f_0$ denotes the inhomogeneous line broadening contributions to the total linewidth. For a two-magnon scattering process mediated by volume and surface defects the latter can be written as:\cite{Nemarich1964}
\begin{equation}
\Delta f_0^{(nmr)}= \Delta f_\text{m-m}F^{(nmr)}+\Delta f_0^0.
\end{equation}
Here, $\Delta f_\text{m-m}$ accounts for the two-magnon scattering process of the MSMs into the spin-wave manifold.\cite{Sparks1961,Nemarich1964,Lam1965} The factor $F^{(nmr)}$ represents the ratio of the linewidth of a particular MSM with respect to the uniform precessing (110)-mode.\cite{Nemarich1964,Lam1965, White1959, White1960} It therefore accounts for the surface sensitivity of the specific mode compared to the (110)-mode. The two-magnon scattering processes can be suppressed if a perfectly polished YIG sphere is used, due to the vanishing ability of the system to transfer linear and angular momentum from and to the lattice.\cite{Nemarich1964} 
The term $\Delta f_0^0$ represents a constant contribution to the linewidth in which all other frequency-independent broadening effects are absorbed. The complete scattering theory used in this letter is presented in Ref.~\onlinecite{Nemarich1964}.

\begin{figure}[t!]%
	\begin{center}%
		\scalebox{1}{\includegraphics[width=\linewidth,clip]{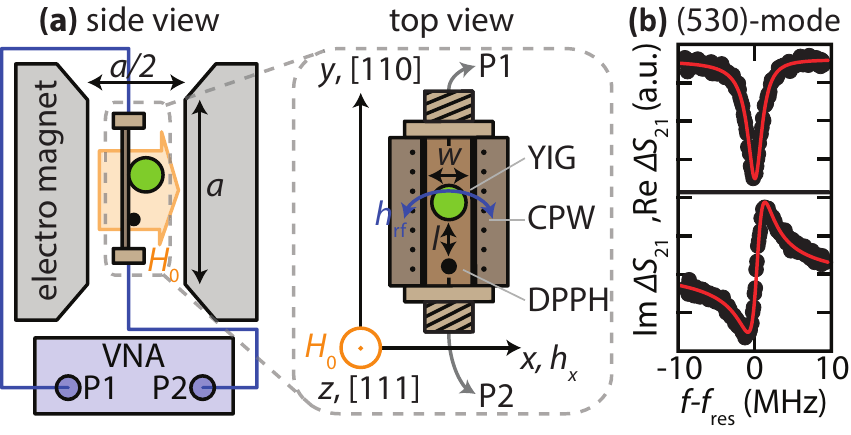}}%
	\end{center}%
	\caption{\label{setup} 
		(a) The CPW with the YIG sphere and the DPPH is positioned in the homogeneous field of an electromagnet. The CPW is connected to port 1 (P1) and port 2 (P2) of a vector network analyzer (VNA). The YIG sphere is placed on top of the center conductor of the CPW with its [111]-axis parallel to the applied magnetic field $H_0$ in $z$-direction. (b)~Typical normalized transmission spectrum of the (530)-mode at $\mu_0H_0=0.8$\,T (symbols) including a fit to Eq.~(\ref{fitfunc}) (lines).}%
\end{figure}%

Fig.~\ref{setup}\,(a) shows a sketch of the measurement setup. The YIG sphere with a diameter of $d=300\,\mu$m is placed in a disk shaped Vespel sample holder (diameter 6\,mm, not shown), which has a centered hole with a diameter of 350\,$\mu$m. The sphere in the sample holder is exposed to a static magnetic field in order to align the easy [111]-direction of the YIG crystal parallel to the field direction. The orientation of the sphere is subsequently fixed using photoresist and the alignment is confirmed by Laue diffraction.

The oriented YIG sphere is placed on a $50\,\Omega$ impedance matched coplanar waveguide (CPW) structure. The sphere is placed in the middle of the $w=300\,\mu$m wide center conductor, with the YIG [110]-axis aligned parallel to the long axis of the center conductor of the CPW. Additionally, a pressed crumb of Diphenylpicrylhydrazyl (DPPH) is glued on the center conductor, where the distance between the YIG sphere and the DPPH is $l \approx 1$\,cm. DPPH is a spin marker with a $g$-factor\cite{Poole1996} of $g_\text{DPPH}=2.0036(3)$. The measurement of its resonance frequency 
\begin{equation}
\label{fdpph}
f_\text{DPPH}=g_\text{DPPH}\frac{\mu_\text{B}}{2 \pi \hbar}\mu_0 H^\text{DPPH}_0
\end{equation} provides an independent magnetic field reference at the sample position, in addition to Hall probe measurements. The static magnetic field calculated from the DPPH resonance frequency is denoted as $H_0^\text{DPPH}$. The stray field originating from the YIG sphere at the location of the DPPH creates a systematic measurement error of $\delta \mu_0 H _\text{stray} \leq 40\,\mu$T, as estimated using a dipole approximation.

For the broadband FMR experiments, the CPW is positioned between the pole shoes of an electromagnet with a maximum field strength of $|\mu_0 H_0| \leq 2.25\,$T. The pole shoe diameter is $a=6$\,cm, while the pole shoe separation is $a/2$, to ensure a sufficient homogeneity of the applied magnetic fields. The measured radial field gradient creates a systematic field measurement error of $\delta \mu_0 H_\text{disp}=0.3$\,mT for $l=1$\,cm displacement from the center axis.

The CPW is connected to port 1 (P1) and port 2 (P2) of a vector network analyzer (VNA) and the complex scattering parameter $S_{21}$ is recorded as a function of $H_0$ and $f\leq26.5$\,GHz. The applied microwave power is -20\,dBm to avoid non-linear effects causing additional line broadening.
The microwave current flowing along the center conductor generates a microwave magnetic field predominately in the $x$-direction at the location of the YIG sphere. This results in an oscillating torque on the magnetization, which is aligned in parallel to the $z$-direction by the external static field ${H}_0$. For $f=f_\text{res}^{(nmr)}$, the excited resonant precession of the magnetization results in an absorption of microwave power. 

In order to eliminate the effect of the frequency dependent background transmission of the CPW, the following measurement protocol is applied: First, $S_{21}$ is measured for fixed $H_0$ in a frequency range $f_\text{DPPH}\pm1\,$GHz. Second, $S_{21}$ is measured for the same frequency range at a slightly larger magnetic field $H_0 + \Delta H_0$, with $\mu_0 \Delta H_0=100\,\text{mT}$. Since for this field no YIG and DPPH resonances are present in the observed frequency range, the latter measurement contains the pure background transmission. Third, the normalized transmission spectra is obtained as $\Delta S_{21}= S_{21}(H_0)/S_{21}(H_0 + \Delta H_0)$, which corrects the magnitude and the phase of the signal. This procedure is repeated for all applied magnetic fields. The transmitted magnitude around the resonance can be expressed as:\cite{Kalarickal2006}
\begin{equation}
\label{fitfunc}
\Delta S_{21}(f)=A+Bf+\frac{Z}{\left(f^{(nmr)}_\text{res}\right) ^2-if^2-if\Delta f^{(nmr)}}.
\end{equation}
Here, $A$ is a complex offset parameter, $B$ is a complex linear background and $Z$ is a complex scaling parameter.\footnote{$A$ and $B$ account for a small drift in $S_{21}$, resulting in an imperfect background correction.} Fig.~\ref{setup}\,(b) exemplary shows the real and imaginary part of $\Delta S_{21}$ for the (530)-mode at $\mu_0 H_0=0.8$\,T. In addition, a fit of Eq.~\ref{fitfunc} to the data is shown, which adequately models the shape of the resonances. 

\begin{figure}[t!]%
	\begin{center}%
		\scalebox{1}{\includegraphics[width=\linewidth,clip]{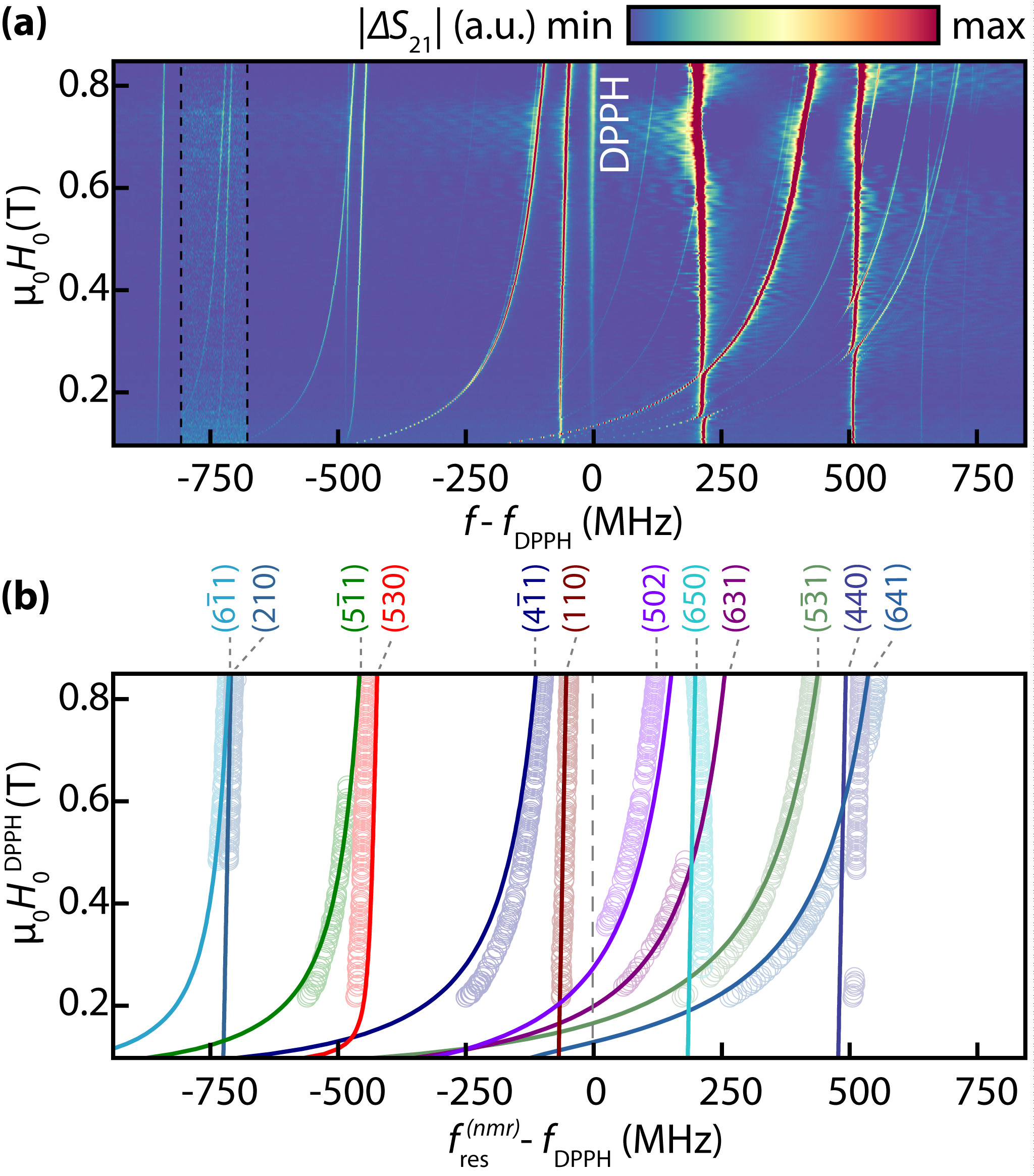}}%
	\end{center}%
	\caption{\label{colormap} 
		(a)~Normalized transmission magnitude $|\Delta S_{21}|$ plotted versus applied magnetic field $\mu_0H_0$ and microwave frequency $f$ relative to the DPPH resonance $f_\text{DPPH}$. The contrast between the dashed lines is stretched for better visibility. (b) Calculated and measured dispersions of various MSMs (lines and open circles, respectively).  
	}%
\end{figure}%

Fig.~\ref{colormap}\,(a) shows the normalized transmitted magnitude $|\Delta S_{21}|$ as a function of $H_0$ and $f-f_\text{DPPH}$ on a linear color-coded scale. The frequency axis is chosen relative to the DPPH resonance frequency, so that all modes with a linear dispersion $f_\text{res}^{(nmr)} \propto H_0$ appear as straight lines, whereas modes with a non-linear dispersion are curved. Note, that the field values displayed on the $y$-axis represent the magnetic field strength measured with the Hall probe.

The different modes appearing in the color plot in Fig.~\ref{colormap}\,(a) can be identified in a straightforward manner. 
At first, all visible resonances are fitted using Eq.~(\ref{fitfunc}) in order to extract $f^{(nmr)}_\text{res}$ and $\Delta f^{(nmr)}$. Furthermore, the DPPH resonance line is identified as straight line at $f-f_\text{DPPH}=0$\,MHz and the resonance fields $H_0^\text{DPPH}$ are calculated using Eq.~(\ref{fdpph}).  

Second, the straight lines at about $f-f_\text{DPPH}\approx-60$\,MHz and $f-f_\text{DPPH} \approx-740$\,MHz are identified as the (110)- and (210)-mode, respectively. A simultaneous fit of the dispersion relations\cite{Fletcher1959a}
\begin{equation}
f_\text{res}^{(110)}= \frac{g_\text{YIG} \mu_\text{B}}{2 \pi \hbar}\mu_0\left(H_0+H_\text{ani}\right)
\end{equation}
and
\begin{equation}
f_{\rm res}^{(210)}= \frac{g_\text{YIG} \mu_\text{B}}{2 \pi \hbar}\mu_0\left(H_0+H_\text{ani}-\frac{2}{15} M_\text{s}\right)
\end{equation}
to the measured values of $f_\text{res}^{(110)}$, $f_\text{res}^{(210)}$ and $\mu_0H_0^\text{DPPH}$
yields $g_\text{YIG}=2.0054(3)$, $\mu_0 M_\text{s}=176.0(4)$\,mT and $\mu_0H_\text{ani}=-2.5(4)$\,mT. The error of $g_\text{YIG}$ is given by the systematic error introduced by the field normalization using $g_\text{DPPH}$. The errors in $\mu_0 H_\text{ani}$ and $\mu_0 M_\text{s}$ are given by $\delta \mu_0 H_\text{disp}+\delta \mu_0 H_\text{stray}$. All values are in good agreement with previously reported material parameters\cite{Dillon1957,Roschmann1983,Hansen1974a,Winkler1981, Roschmann1983,Hansen1974} for YIG ($g_\text{YIG} = 2.005(2)$, $\mu_0 H_\text{ani}=-5.7\,\text{mT}$ and $\mu_0 M_\text{s}=180$\,mT) and, hence, justify the (110)- and (210)-mode assignments.

Third, the complete MSM manifold is computed using the extracted material parameters. The mode numbers of the remaining modes are determined from the characteristic dispersions. 
Fig.~\ref{colormap}\,(b) shows the dispersions of the identified modes as function of $f_\text{res}^{(nmr)}-f_\text{DPPH}$ and $H_0^\text{DPPH}$, with very good agreement of theory (lines) and experiment (circles). Slight deviations between model predictions and data might be attributed to a non-perfect spherical shape of the sample, which would change the boundary conditions for the magnetization dynamic in the YIG spheroid, and thus the dispersion relations.

\begin{figure}[t!] %
	\begin{center}%
		\scalebox{1}{\includegraphics[width=\linewidth,clip]{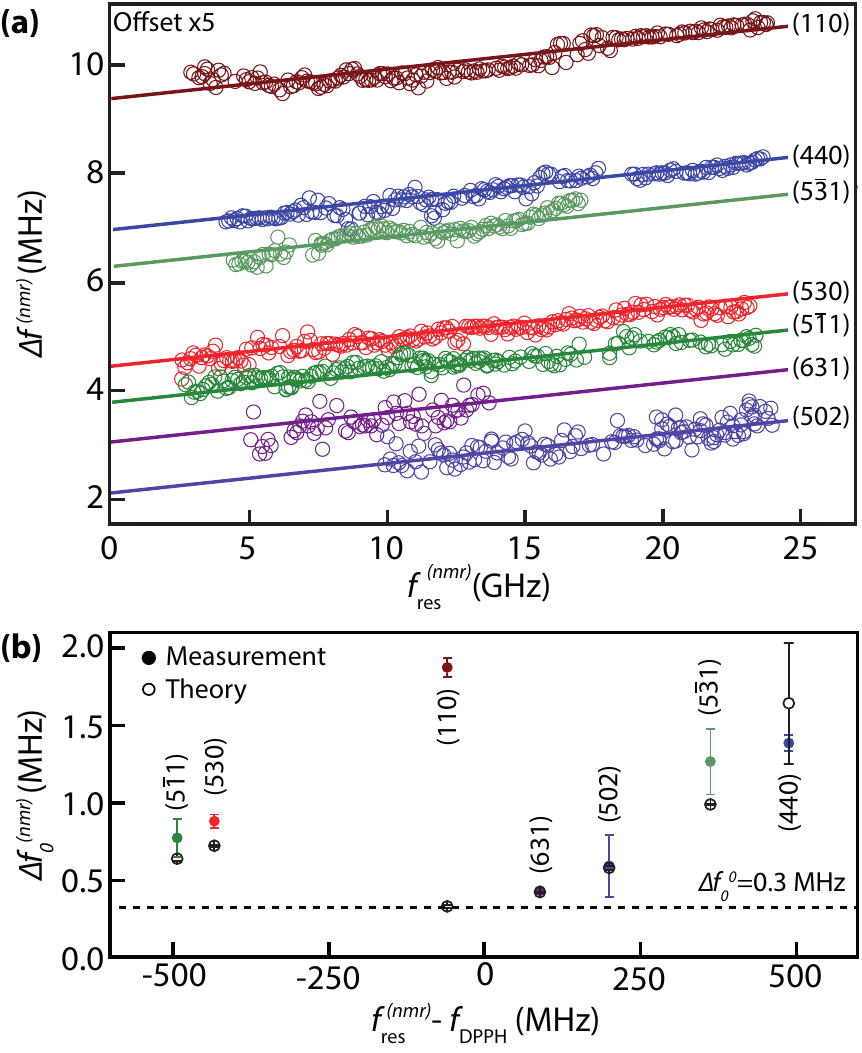}}%
	\end{center}%
	\caption{\label{damping} 
		(a) Linewidth vs. resonance frequency of the measured MSMs. The Gilbert damping of all MSMs is $\alpha=2.7(5)\times 10^{-5}$ as evident from the same slope of all curves. The inhomogeneous line broadening is different for each MSM. Note that the data points are plotted with an offset proportional to the inhomogeneous line broadening. (b) Inhomogeneous line broadening as a function of $f-f_\text{DPPH}$.
	}%
\end{figure}%

In Fig.~\ref{damping}\,(a) the linewidth $\Delta f^{(nmr)}$ of each MSM is plotted versus its resonance frequency $f^{(nmr)}_\text{res}$. The offset $\Delta f_0^{(nmr)}$ is magnified by a factor of 5 to emphasize the differences in the inhomogeneous line broadening. Individual fits of all $\Delta f^{(nmr)}$ to Eq.~(\ref{gilbert_func}) yield identical slopes for all modes within a small scatter, which is also evident from the linewidth data in Fig.~\ref{damping}\,(a). Hence, the Gilbert damping parameter and inhomogeneous line broadening are obtained from a simultaneous fit of Eq.~(\ref{gilbert_func}) to the extracted data points. Here, $\alpha$ is a shared fit parameter for all MSMs, but the inhomogeneous line broadening $\Delta f_0^{(nmr)}$ is fitted separately for each mode. To avoid fitting errors, the linewidths data are disregarded when a mode anti-crossing is observed, since this results in a pronounced increase in linewidth.\cite{Sparks1964} As evident from the solid fit curves in Fig.~\ref{damping}\,(a) the evolution of the linewidth with resonance frequency of all measured MSMs can be well described with a shared Gilbert damping parameter of $\alpha=2.7(5)\times 10^{-5}$, independent of the mode number and the mode intensity. The latter strongly suggests a negligible effect of radiative damping on the measured linewidths.\cite{Schoen2015} The error in $\alpha$ is given by the scatter of $\alpha$ from the independent fits. 
Other groups report Gilbert damping parameters for YIG films\cite{Haertinger2015, DAllivyKelly2013,Hauser2016, Sun2012, Pirro2014,Hahn2014,Dubs2016} larger than $\alpha=6.15\times 10^{-5}$, whereas for bulk YIG\cite{Dubs2016,Roschmann1981,Roschmann1983} values of $\alpha=4\times10^{-5}$ are found. Hence, the Gilbert damping parameter obtained here is the smallest experimental value reported so far. The results are in agreement with the notion, that the Gilbert damping parameter is a bulk property which only depends on intrinsic damping effects. However, the inhomogeneous line broadening is indeed different for the various MSMs.

Fig.~\ref{damping}\,(b) shows the extracted values for the inhomogeneous line broadening (filled dots) as a function of $f^{(nmr)}_\text{res}-f_\text{DPPH}$. The error bars indicate the variation of the inhomogeneous line broadening between global and individual fits. In order to show the approximate position of the modes in comparison to Fig.~\ref{colormap}, the $x$-scale is calculated for a magnetic field strength of $\mu_0H=0.5$\,T. Additionally, the linewidths $\Delta f_0^{(nmr)}$ for all modes are calculated using the two-magnon scattering theory, given in Eq.~(4) of Ref.~\onlinecite{Nemarich1964} (open circles). For the calculations of the linewidths, a pit radius $R=350$\,nm and a constant linewidth contribution of $\Delta f_0^0=30$\,kHz was assumed. Since the calculated $\Delta f_\text{m-m}$ are slightly frequency dependent, the average linewidth values  for the measured field and frequency range are used and the standard deviation is indicated by the error bars of the open symbols. For most MSMs the variation is smaller than 10\,kHz. Nevertheless, the (440)-mode should show a prominent peak in the linewidth measurement at about $f^{(440)}_\text{res}=10$\,GHz in Fig.~\ref{damping}\,(a),\cite{Nemarich1964} which is however not observed in the experimental data. Additionally, the (110)-MSM shows a much larger linewidth than expected from the calculations. In a perfect sphere the (110)-mode is degenerate with the (430)-mode,\cite{Fletcher1959a} but in a real sphere this degeneracy might be lifted. If the difference of the (110)- and (430)-mode frequencies is smaller than the linewidth of the measured resonance, an additional inhomogeneous line broadening is expected. Indeed, a careful analysis of the (110)-MSM line shape reveals a second resonance line in very close vicinity to the (110)-mode, yielding an artificial inhomogeneous line broadening of this mode. Besides these two MSMs, an excellent quantitative agreement between the two-magnon scattering model and experiment is found.

In conclusion, broadband ferromagnetic resonance experiments on magnetostatic modes in a YIG sphere are presented and various magnetostatic modes are identified. The linewidth analysis of the data allows to distinguish between the Gilbert damping and inhomogeneous line broadening. A very small Gilbert damping parameter of $\alpha=2.7(5)\times 10^{-5}$ is found for all MSMs, independent of their mode indices. Furthermore, the inhomogeneous line broadening differs between the various magnetostatic modes, in agreement with the expectations due to two-magnon scattering processes of the magnetostatic modes into the spin-wave manifold.

Financial support from the DFG via SPP 1538 ''Spin Caloric Transport`` (project GO 944/4) is gratefully acknowledged.


\begin{thebibliography}{50}%
	\makeatletter
	\providecommand \@ifxundefined [1]{%
		\@ifx{#1\undefined}
	}%
	\providecommand \@ifnum [1]{%
		\ifnum #1\expandafter \@firstoftwo
		\else \expandafter \@secondoftwo
		\fi
	}%
	\providecommand \@ifx [1]{%
		\ifx #1\expandafter \@firstoftwo
		\else \expandafter \@secondoftwo
		\fi
	}%
	\providecommand \natexlab [1]{#1}%
	\providecommand \enquote  [1]{``#1''}%
	\providecommand \bibnamefont  [1]{#1}%
	\providecommand \bibfnamefont [1]{#1}%
	\providecommand \citenamefont [1]{#1}%
	\providecommand \href@noop [0]{\@secondoftwo}%
	\providecommand \href [0]{\begingroup \@sanitize@url \@href}%
	\providecommand \@href[1]{\@@startlink{#1}\@@href}%
	\providecommand \@@href[1]{\endgroup#1\@@endlink}%
	\providecommand \@sanitize@url [0]{\catcode `\\12\catcode `\$12\catcode
		`\&12\catcode `\#12\catcode `\^12\catcode `\_12\catcode `\%12\relax}%
	\providecommand \@@startlink[1]{}%
	\providecommand \@@endlink[0]{}%
	\providecommand \url  [0]{\begingroup\@sanitize@url \@url }%
	\providecommand \@url [1]{\endgroup\@href {#1}{\urlprefix }}%
	\providecommand \urlprefix  [0]{URL }%
	\providecommand \Eprint [0]{\href }%
	\providecommand \doibase [0]{http://dx.doi.org/}%
	\providecommand \selectlanguage [0]{\@gobble}%
	\providecommand \bibinfo  [0]{\@secondoftwo}%
	\providecommand \bibfield  [0]{\@secondoftwo}%
	\providecommand \translation [1]{[#1]}%
	\providecommand \BibitemOpen [0]{}%
	\providecommand \bibitemStop [0]{}%
	\providecommand \bibitemNoStop [0]{.\EOS\space}%
	\providecommand \EOS [0]{\spacefactor3000\relax}%
	\providecommand \BibitemShut  [1]{\csname bibitem#1\endcsname}%
	\let\auto@bib@innerbib\@empty
	\bibitem [{\citenamefont {Serga}, \citenamefont {Chumak},\ and\ \citenamefont
		{Hillebrands}(2010)}]{Serga2010}%
	\BibitemOpen
	\bibfield  {author} {\bibinfo {author} {\bibfnamefont {A.~A.}\ \bibnamefont
			{Serga}}, \bibinfo {author} {\bibfnamefont {A.~V.}\ \bibnamefont {Chumak}}, \
		and\ \bibinfo {author} {\bibfnamefont {B.}~\bibnamefont {Hillebrands}},\
	}\href {\doibase 10.1088/0022-3727/43/26/264002} {\bibfield  {journal}
	{\bibinfo  {journal} {Journal of Physics D: Applied Physics}\ }\textbf
	{\bibinfo {volume} {43}},\ \bibinfo {pages} {264002} (\bibinfo {year}
	{2010})}\BibitemShut {NoStop}%
\bibitem [{\citenamefont {Chumak}, \citenamefont {Serga},\ and\ \citenamefont
	{Hillebrands}(2014)}]{Chumak2014}%
\BibitemOpen
\bibfield  {author} {\bibinfo {author} {\bibfnamefont {A.~V.}\ \bibnamefont
		{Chumak}}, \bibinfo {author} {\bibfnamefont {A.~A.}\ \bibnamefont {Serga}}, \
	and\ \bibinfo {author} {\bibfnamefont {B.}~\bibnamefont {Hillebrands}},\
}\href {\doibase 10.1038/ncomms5700} {\bibfield  {journal} {\bibinfo
	{journal} {Nature Communications}\ }\textbf {\bibinfo {volume} {5}},\
\bibinfo {pages} {4700} (\bibinfo {year} {2014})}\BibitemShut {NoStop}%
\bibitem [{\citenamefont {Klingler}\ \emph {et~al.}(2015)\citenamefont
	{Klingler}, \citenamefont {Pirro}, \citenamefont {Br{\"{a}}cher},
	\citenamefont {Leven}, \citenamefont {Hillebrands},\ and\ \citenamefont
	{Chumak}}]{Klingler2015}%
\BibitemOpen
\bibfield  {author} {\bibinfo {author} {\bibfnamefont {S.}~\bibnamefont
		{Klingler}}, \bibinfo {author} {\bibfnamefont {P.}~\bibnamefont {Pirro}},
	\bibinfo {author} {\bibfnamefont {T.}~\bibnamefont {Br{\"{a}}cher}}, \bibinfo
	{author} {\bibfnamefont {B.}~\bibnamefont {Leven}}, \bibinfo {author}
	{\bibfnamefont {B.}~\bibnamefont {Hillebrands}}, \ and\ \bibinfo {author}
	{\bibfnamefont {A.~V.}\ \bibnamefont {Chumak}},\ }\href {\doibase
	10.1063/1.4921850} {\bibfield  {journal} {\bibinfo  {journal} {Applied
			Physics Letters}\ }\textbf {\bibinfo {volume} {106}},\ \bibinfo {pages}
	{212406} (\bibinfo {year} {2015})}\BibitemShut {NoStop}%
\bibitem [{\citenamefont {Ganzhorn}\ \emph {et~al.}(2016)\citenamefont
	{Ganzhorn}, \citenamefont {Klingler}, \citenamefont {Wimmer}, \citenamefont
	{Gepr{\"{a}}gs}, \citenamefont {Gross}, \citenamefont {Huebl},\ and\
	\citenamefont {Goennenwein}}]{Ganzhorn2016}%
\BibitemOpen
\bibfield  {author} {\bibinfo {author} {\bibfnamefont {K.}~\bibnamefont
		{Ganzhorn}}, \bibinfo {author} {\bibfnamefont {S.}~\bibnamefont {Klingler}},
	\bibinfo {author} {\bibfnamefont {T.}~\bibnamefont {Wimmer}}, \bibinfo
	{author} {\bibfnamefont {S.}~\bibnamefont {Gepr{\"{a}}gs}}, \bibinfo {author}
	{\bibfnamefont {R.}~\bibnamefont {Gross}}, \bibinfo {author} {\bibfnamefont
		{H.}~\bibnamefont {Huebl}}, \ and\ \bibinfo {author} {\bibfnamefont
		{S.~T.~B.}\ \bibnamefont {Goennenwein}},\ }\href {\doibase 10.1063/1.4958893}
{\bibfield  {journal} {\bibinfo  {journal} {Applied Physics Letters}\
	}\textbf {\bibinfo {volume} {109}},\ \bibinfo {pages} {022405} (\bibinfo
	{year} {2016})}\BibitemShut {NoStop}%
\bibitem [{\citenamefont {Uchida}\ \emph {et~al.}(2008)\citenamefont {Uchida},
	\citenamefont {Takahashi}, \citenamefont {Harii}, \citenamefont {Ieda},
	\citenamefont {Koshibae}, \citenamefont {Ando}, \citenamefont {Maekawa},\
	and\ \citenamefont {Saitoh}}]{Uchida2008}%
\BibitemOpen
\bibfield  {author} {\bibinfo {author} {\bibfnamefont {K.}~\bibnamefont
		{Uchida}}, \bibinfo {author} {\bibfnamefont {S.}~\bibnamefont {Takahashi}},
	\bibinfo {author} {\bibfnamefont {K.}~\bibnamefont {Harii}}, \bibinfo
	{author} {\bibfnamefont {J.}~\bibnamefont {Ieda}}, \bibinfo {author}
	{\bibfnamefont {W.}~\bibnamefont {Koshibae}}, \bibinfo {author}
	{\bibfnamefont {K.}~\bibnamefont {Ando}}, \bibinfo {author} {\bibfnamefont
		{S.}~\bibnamefont {Maekawa}}, \ and\ \bibinfo {author} {\bibfnamefont
		{E.}~\bibnamefont {Saitoh}},\ }\href {\doibase 10.1038/nature07321}
{\bibfield  {journal} {\bibinfo  {journal} {Nature}\ }\textbf {\bibinfo
		{volume} {455}},\ \bibinfo {pages} {778} (\bibinfo {year}
	{2008})}\BibitemShut {NoStop}%
\bibitem [{\citenamefont {Xiao}\ \emph {et~al.}(2010)\citenamefont {Xiao},
	\citenamefont {Bauer}, \citenamefont {Uchida}, \citenamefont {Saitoh},\ and\
	\citenamefont {Maekawa}}]{Xiao2010}%
\BibitemOpen
\bibfield  {author} {\bibinfo {author} {\bibfnamefont {J.}~\bibnamefont
		{Xiao}}, \bibinfo {author} {\bibfnamefont {G.~E.~W.}\ \bibnamefont {Bauer}},
	\bibinfo {author} {\bibfnamefont {K.-C.}\ \bibnamefont {Uchida}}, \bibinfo
	{author} {\bibfnamefont {E.}~\bibnamefont {Saitoh}}, \ and\ \bibinfo {author}
	{\bibfnamefont {S.}~\bibnamefont {Maekawa}},\ }\href {\doibase
	10.1103/PhysRevB.81.214418} {\bibfield  {journal} {\bibinfo  {journal}
		{Physical Review B}\ }\textbf {\bibinfo {volume} {81}},\ \bibinfo {pages}
	{214418} (\bibinfo {year} {2010})}\BibitemShut {NoStop}%
\bibitem [{\citenamefont {Huebl}\ \emph {et~al.}(2013)\citenamefont {Huebl},
	\citenamefont {Zollitsch}, \citenamefont {Lotze}, \citenamefont {Hocke},
	\citenamefont {Greifenstein}, \citenamefont {Marx}, \citenamefont {Gross},\
	and\ \citenamefont {Goennenwein}}]{Huebl2013}%
\BibitemOpen
\bibfield  {author} {\bibinfo {author} {\bibfnamefont {H.}~\bibnamefont
		{Huebl}}, \bibinfo {author} {\bibfnamefont {C.~W.}\ \bibnamefont
		{Zollitsch}}, \bibinfo {author} {\bibfnamefont {J.}~\bibnamefont {Lotze}},
	\bibinfo {author} {\bibfnamefont {F.}~\bibnamefont {Hocke}}, \bibinfo
	{author} {\bibfnamefont {M.}~\bibnamefont {Greifenstein}}, \bibinfo {author}
	{\bibfnamefont {A.}~\bibnamefont {Marx}}, \bibinfo {author} {\bibfnamefont
		{R.}~\bibnamefont {Gross}}, \ and\ \bibinfo {author} {\bibfnamefont
		{S.~T.~B.}\ \bibnamefont {Goennenwein}},\ }\href {\doibase
	10.1103/PhysRevLett.111.127003} {\bibfield  {journal} {\bibinfo  {journal}
		{Physical Review Letters}\ }\textbf {\bibinfo {volume} {111}},\ \bibinfo
	{pages} {127003} (\bibinfo {year} {2013})}\BibitemShut {NoStop}%
\bibitem [{\citenamefont {Soykal}\ and\ \citenamefont
	{Flatt{\'{e}}}(2010{\natexlab{a}})}]{Soykal2010}%
\BibitemOpen
\bibfield  {author} {\bibinfo {author} {\bibfnamefont {{\"{O}}.~O.}\
		\bibnamefont {Soykal}}\ and\ \bibinfo {author} {\bibfnamefont {M.~E.}\
		\bibnamefont {Flatt{\'{e}}}},\ }\href {\doibase
	10.1103/PhysRevLett.104.077202} {\bibfield  {journal} {\bibinfo  {journal}
		{Physical Review Letters}\ }\textbf {\bibinfo {volume} {104}},\ \bibinfo
	{pages} {077202} (\bibinfo {year} {2010}{\natexlab{a}})}\BibitemShut
{NoStop}%
\bibitem [{\citenamefont {Soykal}\ and\ \citenamefont
	{Flatt{\'{e}}}(2010{\natexlab{b}})}]{Soykal2010a}%
\BibitemOpen
\bibfield  {author} {\bibinfo {author} {\bibfnamefont {{\"{O}}.~O.}\
		\bibnamefont {Soykal}}\ and\ \bibinfo {author} {\bibfnamefont {M.~E.}\
		\bibnamefont {Flatt{\'{e}}}},\ }\href {\doibase 10.1103/PhysRevB.82.104413}
{\bibfield  {journal} {\bibinfo  {journal} {Physical Review B}\ }\textbf
	{\bibinfo {volume} {82}},\ \bibinfo {pages} {104413} (\bibinfo {year}
	{2010}{\natexlab{b}})}\BibitemShut {NoStop}%
\bibitem [{\citenamefont {Imamoglu}(2009)}]{Imamoglu2009}%
\BibitemOpen
\bibfield  {author} {\bibinfo {author} {\bibfnamefont {A.}~\bibnamefont
		{Imamoglu}},\ }\href {\doibase 10.1103/PhysRevLett.102.083602} {\bibfield
	{journal} {\bibinfo  {journal} {Physical Review Letters}\ }\textbf {\bibinfo
		{volume} {102}},\ \bibinfo {pages} {083602} (\bibinfo {year}
	{2009})}\BibitemShut {NoStop}%
\bibitem [{\citenamefont {Wesenberg}\ \emph {et~al.}(2009)\citenamefont
	{Wesenberg}, \citenamefont {Ardavan}, \citenamefont {Briggs}, \citenamefont
	{Morton}, \citenamefont {Schoelkopf}, \citenamefont {Schuster},\ and\
	\citenamefont {M{\o}lmer}}]{Wesenberg2009}%
\BibitemOpen
\bibfield  {author} {\bibinfo {author} {\bibfnamefont {J.~H.}\ \bibnamefont
		{Wesenberg}}, \bibinfo {author} {\bibfnamefont {A.}~\bibnamefont {Ardavan}},
	\bibinfo {author} {\bibfnamefont {G.~A.~D.}\ \bibnamefont {Briggs}}, \bibinfo
	{author} {\bibfnamefont {J.~J.~L.}\ \bibnamefont {Morton}}, \bibinfo {author}
	{\bibfnamefont {R.~J.}\ \bibnamefont {Schoelkopf}}, \bibinfo {author}
	{\bibfnamefont {D.~I.}\ \bibnamefont {Schuster}}, \ and\ \bibinfo {author}
	{\bibfnamefont {K.}~\bibnamefont {M{\o}lmer}},\ }\href {\doibase
	10.1103/PhysRevLett.103.070502} {\bibfield  {journal} {\bibinfo  {journal}
		{Physical Review Letters}\ }\textbf {\bibinfo {volume} {103}},\ \bibinfo
	{pages} {070502} (\bibinfo {year} {2009})}\BibitemShut {NoStop}%
\bibitem [{\citenamefont {Tabuchi}\ \emph {et~al.}(2015)\citenamefont
	{Tabuchi}, \citenamefont {Ishino}, \citenamefont {Noguchi}, \citenamefont
	{Ishikawa}, \citenamefont {Yamazaki}, \citenamefont {Usami},\ and\
	\citenamefont {Nakamura}}]{Tabuchi2015}%
\BibitemOpen
\bibfield  {author} {\bibinfo {author} {\bibfnamefont {Y.}~\bibnamefont
		{Tabuchi}}, \bibinfo {author} {\bibfnamefont {S.}~\bibnamefont {Ishino}},
	\bibinfo {author} {\bibfnamefont {A.}~\bibnamefont {Noguchi}}, \bibinfo
	{author} {\bibfnamefont {T.}~\bibnamefont {Ishikawa}}, \bibinfo {author}
	{\bibfnamefont {R.}~\bibnamefont {Yamazaki}}, \bibinfo {author}
	{\bibfnamefont {K.}~\bibnamefont {Usami}}, \ and\ \bibinfo {author}
	{\bibfnamefont {Y.}~\bibnamefont {Nakamura}},\ }\href {\doibase
	10.1126/science.aaa3693} {\bibfield  {journal} {\bibinfo  {journal}
		{Science}\ }\textbf {\bibinfo {volume} {349}},\ \bibinfo {pages} {405}
	(\bibinfo {year} {2015})}\BibitemShut {NoStop}%
\bibitem [{\citenamefont {Maier-Flaig}\ \emph {et~al.}(2016)\citenamefont
	{Maier-Flaig}, \citenamefont {Harder}, \citenamefont {Gross}, \citenamefont
	{Huebl},\ and\ \citenamefont {Goennenwein}}]{Maier-Flaig2016}%
\BibitemOpen
\bibfield  {author} {\bibinfo {author} {\bibfnamefont {H.}~\bibnamefont
		{Maier-Flaig}}, \bibinfo {author} {\bibfnamefont {M.}~\bibnamefont {Harder}},
	\bibinfo {author} {\bibfnamefont {R.}~\bibnamefont {Gross}}, \bibinfo
	{author} {\bibfnamefont {H.}~\bibnamefont {Huebl}}, \ and\ \bibinfo {author}
	{\bibfnamefont {S.~T.~B.}\ \bibnamefont {Goennenwein}},\ }\href {\doibase
	10.1103/PhysRevB.94.054433} {\bibfield  {journal} {\bibinfo  {journal}
		{Physical Review B}\ }\textbf {\bibinfo {volume} {94}},\ \bibinfo {pages}
	{054433} (\bibinfo {year} {2016})}\BibitemShut {NoStop}%
\bibitem [{\citenamefont {Klingler}\ \emph {et~al.}(2016)\citenamefont
	{Klingler}, \citenamefont {Maier-Flaig}, \citenamefont {Gross}, \citenamefont
	{Hu}, \citenamefont {Huebl}, \citenamefont {Goennenwein},\ and\ \citenamefont
	{Weiler}}]{Klingler2016}%
\BibitemOpen
\bibfield  {author} {\bibinfo {author} {\bibfnamefont {S.}~\bibnamefont
		{Klingler}}, \bibinfo {author} {\bibfnamefont {H.}~\bibnamefont
		{Maier-Flaig}}, \bibinfo {author} {\bibfnamefont {R.}~\bibnamefont {Gross}},
	\bibinfo {author} {\bibfnamefont {C.-M.}\ \bibnamefont {Hu}}, \bibinfo
	{author} {\bibfnamefont {H.}~\bibnamefont {Huebl}}, \bibinfo {author}
	{\bibfnamefont {S.~T.~B.}\ \bibnamefont {Goennenwein}}, \ and\ \bibinfo
	{author} {\bibfnamefont {M.}~\bibnamefont {Weiler}},\ }\href {\doibase
	10.1063/1.4961052} {\bibfield  {journal} {\bibinfo  {journal} {Applied
			Physics Letters}\ }\textbf {\bibinfo {volume} {109}},\ \bibinfo {pages}
	{072402} (\bibinfo {year} {2016})}\BibitemShut {NoStop}%
\bibitem [{\citenamefont {Hisatomi}\ \emph {et~al.}(2016)\citenamefont
	{Hisatomi}, \citenamefont {Osada}, \citenamefont {Tabuchi}, \citenamefont
	{Ishikawa}, \citenamefont {Noguchi}, \citenamefont {Yamazaki}, \citenamefont
	{Usami},\ and\ \citenamefont {Nakamura}}]{Hisatomi2016}%
\BibitemOpen
\bibfield  {author} {\bibinfo {author} {\bibfnamefont {R.}~\bibnamefont
		{Hisatomi}}, \bibinfo {author} {\bibfnamefont {A.}~\bibnamefont {Osada}},
	\bibinfo {author} {\bibfnamefont {Y.}~\bibnamefont {Tabuchi}}, \bibinfo
	{author} {\bibfnamefont {T.}~\bibnamefont {Ishikawa}}, \bibinfo {author}
	{\bibfnamefont {A.}~\bibnamefont {Noguchi}}, \bibinfo {author} {\bibfnamefont
		{R.}~\bibnamefont {Yamazaki}}, \bibinfo {author} {\bibfnamefont
		{K.}~\bibnamefont {Usami}}, \ and\ \bibinfo {author} {\bibfnamefont
		{Y.}~\bibnamefont {Nakamura}},\ }\href {\doibase 10.1103/PhysRevB.93.174427}
{\bibfield  {journal} {\bibinfo  {journal} {Physical Review B}\ }\textbf
	{\bibinfo {volume} {93}},\ \bibinfo {pages} {174427} (\bibinfo {year}
	{2016})}\BibitemShut {NoStop}%
\bibitem [{\citenamefont {Zhang}\ \emph {et~al.}(2016)\citenamefont {Zhang},
	\citenamefont {Zhu}, \citenamefont {Zou},\ and\ \citenamefont
	{Tang}}]{Zhang2016}%
\BibitemOpen
\bibfield  {author} {\bibinfo {author} {\bibfnamefont {X.}~\bibnamefont
		{Zhang}}, \bibinfo {author} {\bibfnamefont {N.}~\bibnamefont {Zhu}}, \bibinfo
	{author} {\bibfnamefont {C.-L.}\ \bibnamefont {Zou}}, \ and\ \bibinfo
	{author} {\bibfnamefont {H.~X.}\ \bibnamefont {Tang}},\ }\href {\doibase
	10.1103/PhysRevLett.117.123605} {\bibfield  {journal} {\bibinfo  {journal}
		{Physical Review Letters}\ }\textbf {\bibinfo {volume} {117}},\ \bibinfo
	{pages} {123605} (\bibinfo {year} {2016})}\BibitemShut {NoStop}%
\bibitem [{\citenamefont {Walker}(1957)}]{Walker1957}%
\BibitemOpen
\bibfield  {author} {\bibinfo {author} {\bibfnamefont {L.~R.}\ \bibnamefont
		{Walker}},\ }\href {\doibase 10.1103/PhysRev.105.390} {\bibfield  {journal}
	{\bibinfo  {journal} {Physical Review}\ }\textbf {\bibinfo {volume} {105}},\
	\bibinfo {pages} {390} (\bibinfo {year} {1957})}\BibitemShut {NoStop}%
\bibitem [{\citenamefont {Fletcher}\ and\ \citenamefont
	{Bell}(1959)}]{Fletcher1959a}%
\BibitemOpen
\bibfield  {author} {\bibinfo {author} {\bibfnamefont {P.~C.}\ \bibnamefont
		{Fletcher}}\ and\ \bibinfo {author} {\bibfnamefont {R.~O.}\ \bibnamefont
		{Bell}},\ }\href {\doibase 10.1063/1.1735216} {\bibfield  {journal} {\bibinfo
		{journal} {Journal of Applied Physics}\ }\textbf {\bibinfo {volume} {30}},\
	\bibinfo {pages} {687} (\bibinfo {year} {1959})}\BibitemShut {NoStop}%
\bibitem [{\citenamefont {R{\"{o}}schmann}\ and\ \citenamefont
	{D{\"{o}}tsch}(1977)}]{Roschmann1977}%
\BibitemOpen
\bibfield  {author} {\bibinfo {author} {\bibfnamefont {P.}~\bibnamefont
		{R{\"{o}}schmann}}\ and\ \bibinfo {author} {\bibfnamefont {H.}~\bibnamefont
		{D{\"{o}}tsch}},\ }\href {\doibase 10.1002/pssb.2220820102} {\bibfield
	{journal} {\bibinfo  {journal} {Physica Status Solidi (b)}\ }\textbf
	{\bibinfo {volume} {82}},\ \bibinfo {pages} {11} (\bibinfo {year}
	{1977})}\BibitemShut {NoStop}%
\bibitem [{\citenamefont {R{\"{o}}schmann}(1975)}]{Roschmann1975}%
\BibitemOpen
\bibfield  {author} {\bibinfo {author} {\bibfnamefont {P.}~\bibnamefont
		{R{\"{o}}schmann}},\ }\href {\doibase 10.1109/TMAG.1975.1058912} {\bibfield
	{journal} {\bibinfo  {journal} {IEEE Transactions on Magnetics}\ }\textbf
	{\bibinfo {volume} {11}},\ \bibinfo {pages} {1247} (\bibinfo {year}
	{1975})}\BibitemShut {NoStop}%
\bibitem [{\citenamefont {Nemarich}(1964)}]{Nemarich1964}%
\BibitemOpen
\bibfield  {author} {\bibinfo {author} {\bibfnamefont {J.}~\bibnamefont
		{Nemarich}},\ }\href {\doibase 10.1103/PhysRev.136.A1657} {\bibfield
	{journal} {\bibinfo  {journal} {Physical Review}\ }\textbf {\bibinfo {volume}
		{136}},\ \bibinfo {pages} {A1657} (\bibinfo {year} {1964})}\BibitemShut
{NoStop}%
\bibitem [{\citenamefont {Lam}(1965)}]{Lam1965}%
\BibitemOpen
\bibfield  {author} {\bibinfo {author} {\bibfnamefont {Y.}~\bibnamefont
		{Lam}},\ }\href {\doibase 10.1016/0038-1101(65)90155-3} {\bibfield  {journal}
	{\bibinfo  {journal} {Solid-State Electronics}\ }\textbf {\bibinfo {volume}
		{8}},\ \bibinfo {pages} {923} (\bibinfo {year} {1965})}\BibitemShut {NoStop}%
\bibitem [{Note1()}]{Note1}%
\BibitemOpen
\bibinfo {note} {Here, inhomogeneous refers to all non-Gilbert like damping
	mechanisms, while the lineshape remains Lorentzian.}\BibitemShut {Stop}%
\bibitem [{\citenamefont {H{\"{o}}lzer}(1992)}]{Holzer1992}%
\BibitemOpen
\bibfield  {author} {\bibinfo {author} {\bibfnamefont {D.}~\bibnamefont
		{H{\"{o}}lzer}},\ }\href {\doibase 10.1016/0375-9601(92)90390-8} {\bibfield
	{journal} {\bibinfo  {journal} {Physics Letters A}\ }\textbf {\bibinfo
		{volume} {170}},\ \bibinfo {pages} {45} (\bibinfo {year} {1992})}\BibitemShut
{NoStop}%
\bibitem [{\citenamefont {Stancil}\ and\ \citenamefont
	{Prabhakar}(2009)}]{Stancil2009}%
\BibitemOpen
\bibfield  {author} {\bibinfo {author} {\bibfnamefont {D.~D.}\ \bibnamefont
		{Stancil}}\ and\ \bibinfo {author} {\bibfnamefont {A.}~\bibnamefont
		{Prabhakar}},\ }\href {\doibase 10.1007/978-0-387-77865-5} {\emph {\bibinfo
		{title} {{Spin Waves}}}},\ \bibinfo {edition} {1st}\ ed.\ (\bibinfo
{publisher} {Springer US},\ \bibinfo {address} {Boston, MA},\ \bibinfo {year}
{2009})\BibitemShut {NoStop}%
\bibitem [{\citenamefont {Landau}\ and\ \citenamefont
	{Lifshitz}(1935)}]{Landau1935}%
\BibitemOpen
\bibfield  {author} {\bibinfo {author} {\bibfnamefont {L.~D.}\ \bibnamefont
		{Landau}}\ and\ \bibinfo {author} {\bibfnamefont {E.}~\bibnamefont
		{Lifshitz}},\ }\href
{http://ujp.bitp.kiev.ua/files/journals/53/si/53SI06p.pdf
	http://www.ujp.bitp.kiev.ua/files/journals/53/si/53SI06p.pdf} {\bibfield
	{journal} {\bibinfo  {journal} {Phys. Z. Sowjetunion}\ }\textbf {\bibinfo
		{volume} {5}},\ \bibinfo {pages} {153} (\bibinfo {year} {1935})}\BibitemShut
{NoStop}%
\bibitem [{\citenamefont {Gilbert}(2004)}]{Gilbert2004}%
\BibitemOpen
\bibfield  {author} {\bibinfo {author} {\bibfnamefont {T.~L.}\ \bibnamefont
		{Gilbert}},\ }\href {\doibase 10.1109/TMAG.2004.836740} {\bibfield  {journal}
	{\bibinfo  {journal} {IEEE Transactions on Magnetics}\ }\textbf {\bibinfo
		{volume} {40}},\ \bibinfo {pages} {3443} (\bibinfo {year}
	{2004})}\BibitemShut {NoStop}%
\bibitem [{\citenamefont {Wiese}(1991)}]{Wiese1991}%
\BibitemOpen
\bibfield  {author} {\bibinfo {author} {\bibfnamefont {G.}~\bibnamefont
		{Wiese}},\ }\href {\doibase 10.1007/BF01357194} {\bibfield  {journal}
	{\bibinfo  {journal} {Zeitschrift f{\"{u}}r Physik B - Condensed Matter}\
	}\textbf {\bibinfo {volume} {82}},\ \bibinfo {pages} {453} (\bibinfo {year}
	{1991})}\BibitemShut {NoStop}%
\bibitem [{\citenamefont {Fletcher}, \citenamefont {Solt},\ and\ \citenamefont
	{Bell}(1959)}]{Fletcher1959}%
\BibitemOpen
\bibfield  {author} {\bibinfo {author} {\bibfnamefont {P.}~\bibnamefont
		{Fletcher}}, \bibinfo {author} {\bibfnamefont {I.~H.}\ \bibnamefont {Solt}},
	\ and\ \bibinfo {author} {\bibfnamefont {R.}~\bibnamefont {Bell}},\ }\href
{\doibase 10.1103/PhysRev.114.739} {\bibfield  {journal} {\bibinfo  {journal}
		{Physical Review}\ }\textbf {\bibinfo {volume} {114}},\ \bibinfo {pages}
	{739} (\bibinfo {year} {1959})}\BibitemShut {NoStop}%
\bibitem [{\citenamefont {Kalarickal}\ \emph {et~al.}(2006)\citenamefont
	{Kalarickal}, \citenamefont {Krivosik}, \citenamefont {Wu}, \citenamefont
	{Patton}, \citenamefont {Schneider}, \citenamefont {Kabos}, \citenamefont
	{Silva},\ and\ \citenamefont {Nibarger}}]{Kalarickal2006}%
\BibitemOpen
\bibfield  {author} {\bibinfo {author} {\bibfnamefont {S.~S.}\ \bibnamefont
		{Kalarickal}}, \bibinfo {author} {\bibfnamefont {P.}~\bibnamefont
		{Krivosik}}, \bibinfo {author} {\bibfnamefont {M.}~\bibnamefont {Wu}},
	\bibinfo {author} {\bibfnamefont {C.~E.}\ \bibnamefont {Patton}}, \bibinfo
	{author} {\bibfnamefont {M.~L.}\ \bibnamefont {Schneider}}, \bibinfo {author}
	{\bibfnamefont {P.}~\bibnamefont {Kabos}}, \bibinfo {author} {\bibfnamefont
		{T.~J.}\ \bibnamefont {Silva}}, \ and\ \bibinfo {author} {\bibfnamefont
		{J.~P.}\ \bibnamefont {Nibarger}},\ }\href {\doibase 10.1063/1.2197087}
{\bibfield  {journal} {\bibinfo  {journal} {Journal of Applied Physics}\
	}\textbf {\bibinfo {volume} {99}},\ \bibinfo {pages} {093909} (\bibinfo
	{year} {2006})}\BibitemShut {NoStop}%
\bibitem [{\citenamefont {Sparks}, \citenamefont {Loudon},\ and\ \citenamefont
	{Kittel}(1961)}]{Sparks1961}%
\BibitemOpen
\bibfield  {author} {\bibinfo {author} {\bibfnamefont {M.}~\bibnamefont
		{Sparks}}, \bibinfo {author} {\bibfnamefont {R.}~\bibnamefont {Loudon}}, \
	and\ \bibinfo {author} {\bibfnamefont {C.}~\bibnamefont {Kittel}},\ }\href
{\doibase 10.1103/PhysRev.122.791} {\bibfield  {journal} {\bibinfo  {journal}
		{Physical Review}\ }\textbf {\bibinfo {volume} {122}},\ \bibinfo {pages}
	{791} (\bibinfo {year} {1961})}\BibitemShut {NoStop}%
\bibitem [{\citenamefont {White}(1959)}]{White1959}%
\BibitemOpen
\bibfield  {author} {\bibinfo {author} {\bibfnamefont {R.~L.}\ \bibnamefont
		{White}},\ }\href {\doibase 10.1063/1.2185877} {\bibfield  {journal}
	{\bibinfo  {journal} {Journal of Applied Physics}\ }\textbf {\bibinfo
		{volume} {30}},\ \bibinfo {pages} {S182} (\bibinfo {year}
	{1959})}\BibitemShut {NoStop}%
\bibitem [{\citenamefont {White}(1960)}]{White1960}%
\BibitemOpen
\bibfield  {author} {\bibinfo {author} {\bibfnamefont {R.~L.}\ \bibnamefont
		{White}},\ }\href {\doibase 10.1063/1.1984615} {\bibfield  {journal}
	{\bibinfo  {journal} {Journal of Applied Physics}\ }\textbf {\bibinfo
		{volume} {31}},\ \bibinfo {pages} {S86} (\bibinfo {year} {1960})}\BibitemShut
{NoStop}%
\bibitem [{\citenamefont {Poole}(1996)}]{Poole1996}%
\BibitemOpen
\bibfield  {author} {\bibinfo {author} {\bibfnamefont {C.~P.}\ \bibnamefont
		{Poole}},\ }\href
{https://books.google.de/books?hl=de{\&}lr={\&}id=P-4PIoi7Z7IC{\&}oi=fnd{\&}pg=PR22{\&}dq=dpph+g+factor+poole{\&}ots=XnrjqcFM9J{\&}sig=dNucgr0Pm3zZDpj2cFFCuZcSWRg{\#}v=snippet{\&}q=2.0036{\&}f=false
	http://www.amazon.com/Electron-Spin-Resonance-Comprehensive-Experimental/dp/0486694445}
{\emph {\bibinfo {title} {{Electron Spin Resonance}}}},\ \bibinfo {edition}
{2nd}\ ed.\ (\bibinfo  {publisher} {Dover Publications},\ \bibinfo {address}
{Mineola, New York},\ \bibinfo {year} {1996})\BibitemShut {NoStop}%
\bibitem [{Note2()}]{Note2}%
\BibitemOpen
\bibinfo {note} {$A$ and $B$ account for a small drift in $S_{21}$, resulting
	in an imperfect background correction.}\BibitemShut {Stop}%
\bibitem [{\citenamefont {Dillon}(1957)}]{Dillon1957}%
\BibitemOpen
\bibfield  {author} {\bibinfo {author} {\bibfnamefont {J.~F.}\ \bibnamefont
		{Dillon}},\ }\href@noop {} {\bibfield  {journal} {\bibinfo  {journal}
		{Physical Review}\ }\textbf {\bibinfo {volume} {105}},\ \bibinfo {pages}
	{759} (\bibinfo {year} {1957})}\BibitemShut {NoStop}%
\bibitem [{\citenamefont {R{\"{o}}schmann}\ and\ \citenamefont
	{Tolksdorf}(1983)}]{Roschmann1983}%
\BibitemOpen
\bibfield  {author} {\bibinfo {author} {\bibfnamefont {P.}~\bibnamefont
		{R{\"{o}}schmann}}\ and\ \bibinfo {author} {\bibfnamefont {W.}~\bibnamefont
		{Tolksdorf}},\ }\href {\doibase 10.1016/0025-5408(83)90137-X} {\bibfield
	{journal} {\bibinfo  {journal} {Materials Research Bulletin}\ }\textbf
	{\bibinfo {volume} {18}},\ \bibinfo {pages} {449} (\bibinfo {year}
	{1983})}\BibitemShut {NoStop}%
\bibitem [{\citenamefont {Hansen}(1974)}]{Hansen1974a}%
\BibitemOpen
\bibfield  {author} {\bibinfo {author} {\bibfnamefont {P.}~\bibnamefont
		{Hansen}},\ }\href {\doibase 10.1063/1.1663830} {\bibfield  {journal}
	{\bibinfo  {journal} {Journal of Applied Physics}\ }\textbf {\bibinfo
		{volume} {45}},\ \bibinfo {pages} {3638} (\bibinfo {year}
	{1974})}\BibitemShut {NoStop}%
\bibitem [{\citenamefont {Winkler}(1981)}]{Winkler1981}%
\BibitemOpen
\bibfield  {author} {\bibinfo {author} {\bibfnamefont {G.}~\bibnamefont
		{Winkler}},\ }\href@noop {} {\emph {\bibinfo {title} {{Magnetic Garnets}}}},\
\bibinfo {edition} {5th}\ ed.\ (\bibinfo  {publisher} {Vieweg},\ \bibinfo
{address} {Braunschweig, Wiesbaden},\ \bibinfo {year} {1981})\BibitemShut
{NoStop}%
\bibitem [{\citenamefont {Hansen}, \citenamefont {R{\"{o}}schmann},\ and\
	\citenamefont {Tolksdorf}(1974)}]{Hansen1974}%
\BibitemOpen
\bibfield  {author} {\bibinfo {author} {\bibfnamefont {P.}~\bibnamefont
		{Hansen}}, \bibinfo {author} {\bibfnamefont {P.}~\bibnamefont
		{R{\"{o}}schmann}}, \ and\ \bibinfo {author} {\bibfnamefont {W.}~\bibnamefont
		{Tolksdorf}},\ }\href {\doibase 10.1063/1.1663657} {\bibfield  {journal}
	{\bibinfo  {journal} {Journal of Applied Physics}\ }\textbf {\bibinfo
		{volume} {45}},\ \bibinfo {pages} {2728} (\bibinfo {year}
	{1974})}\BibitemShut {NoStop}%
\bibitem [{\citenamefont {Sparks}(1964)}]{Sparks1964}%
\BibitemOpen
\bibfield  {author} {\bibinfo {author} {\bibfnamefont {M.}~\bibnamefont
		{Sparks}},\ }\href@noop {} {\emph {\bibinfo {title}
		{{Ferromagnetic-Relaxation Theory}}}}\ (\bibinfo  {publisher} {McGraw-Hill},\
\bibinfo {year} {1964})\BibitemShut {NoStop}%
\bibitem [{\citenamefont {Schoen}\ \emph {et~al.}(2015)\citenamefont {Schoen},
	\citenamefont {Shaw}, \citenamefont {Nembach}, \citenamefont {Weiler},\ and\
	\citenamefont {Silva}}]{Schoen2015}%
\BibitemOpen
\bibfield  {author} {\bibinfo {author} {\bibfnamefont {M.~A.~W.}\
		\bibnamefont {Schoen}}, \bibinfo {author} {\bibfnamefont {J.~M.}\
		\bibnamefont {Shaw}}, \bibinfo {author} {\bibfnamefont {H.~T.}\ \bibnamefont
		{Nembach}}, \bibinfo {author} {\bibfnamefont {M.}~\bibnamefont {Weiler}}, \
	and\ \bibinfo {author} {\bibfnamefont {T.~J.}\ \bibnamefont {Silva}},\ }\href
{\doibase 10.1103/PhysRevB.92.184417} {\bibfield  {journal} {\bibinfo
		{journal} {Physical Review B}\ }\textbf {\bibinfo {volume} {92}},\ \bibinfo
	{pages} {184417} (\bibinfo {year} {2015})}\BibitemShut {NoStop}%
\bibitem [{\citenamefont {Haertinger}\ \emph {et~al.}(2015)\citenamefont
	{Haertinger}, \citenamefont {Back}, \citenamefont {Lotze}, \citenamefont
	{Weiler}, \citenamefont {Gepr{\"{a}}gs}, \citenamefont {Huebl}, \citenamefont
	{Goennenwein},\ and\ \citenamefont {Woltersdorf}}]{Haertinger2015}%
\BibitemOpen
\bibfield  {author} {\bibinfo {author} {\bibfnamefont {M.}~\bibnamefont
		{Haertinger}}, \bibinfo {author} {\bibfnamefont {C.~H.}\ \bibnamefont
		{Back}}, \bibinfo {author} {\bibfnamefont {J.}~\bibnamefont {Lotze}},
	\bibinfo {author} {\bibfnamefont {M.}~\bibnamefont {Weiler}}, \bibinfo
	{author} {\bibfnamefont {S.}~\bibnamefont {Gepr{\"{a}}gs}}, \bibinfo {author}
	{\bibfnamefont {H.}~\bibnamefont {Huebl}}, \bibinfo {author} {\bibfnamefont
		{S.~T.~B.}\ \bibnamefont {Goennenwein}}, \ and\ \bibinfo {author}
	{\bibfnamefont {G.}~\bibnamefont {Woltersdorf}},\ }\href {\doibase
	10.1103/PhysRevB.92.054437} {\bibfield  {journal} {\bibinfo  {journal}
		{Physical Review B}\ }\textbf {\bibinfo {volume} {92}},\ \bibinfo {pages}
	{054437} (\bibinfo {year} {2015})}\BibitemShut {NoStop}%
\bibitem [{\citenamefont {d'Allivy Kelly}\ \emph {et~al.}(2013)\citenamefont
	{d'Allivy Kelly}, \citenamefont {Anane}, \citenamefont {Bernard},
	\citenamefont {{Ben Youssef}}, \citenamefont {Hahn}, \citenamefont
	{Molpeceres}, \citenamefont {Carrétéro}, \citenamefont {Jacquet},
	\citenamefont {Deranlot}, \citenamefont {Bortolotti}, \citenamefont
	{Lebourgeois}, \citenamefont {Mage}, \citenamefont {de~Loubens},
	\citenamefont {Klein}, \citenamefont {Cros},\ and\ \citenamefont
	{Fert}}]{DAllivyKelly2013}%
\BibitemOpen
\bibfield  {author} {\bibinfo {author} {\bibfnamefont {O.}~\bibnamefont
		{d'Allivy Kelly}}, \bibinfo {author} {\bibfnamefont {A.}~\bibnamefont
		{Anane}}, \bibinfo {author} {\bibfnamefont {R.}~\bibnamefont {Bernard}},
	\bibinfo {author} {\bibfnamefont {J.}~\bibnamefont {{Ben Youssef}}}, \bibinfo
	{author} {\bibfnamefont {C.}~\bibnamefont {Hahn}}, \bibinfo {author}
	{\bibfnamefont {A.~H.}\ \bibnamefont {Molpeceres}}, \bibinfo {author}
	{\bibfnamefont {C.}~\bibnamefont {Carrétéro}}, \bibinfo {author}
	{\bibfnamefont {E.}~\bibnamefont {Jacquet}}, \bibinfo {author} {\bibfnamefont
		{C.}~\bibnamefont {Deranlot}}, \bibinfo {author} {\bibfnamefont
		{P.}~\bibnamefont {Bortolotti}}, \bibinfo {author} {\bibfnamefont
		{R.}~\bibnamefont {Lebourgeois}}, \bibinfo {author} {\bibfnamefont {J.-C.}\
		\bibnamefont {Mage}}, \bibinfo {author} {\bibfnamefont {G.}~\bibnamefont
		{de~Loubens}}, \bibinfo {author} {\bibfnamefont {O.}~\bibnamefont {Klein}},
	\bibinfo {author} {\bibfnamefont {V.}~\bibnamefont {Cros}}, \ and\ \bibinfo
	{author} {\bibfnamefont {A.}~\bibnamefont {Fert}},\ }\href {\doibase
	10.1063/1.4819157} {\bibfield  {journal} {\bibinfo  {journal} {Applied
			Physics Letters}\ }\textbf {\bibinfo {volume} {103}},\ \bibinfo {pages}
	{082408} (\bibinfo {year} {2013})}\BibitemShut {NoStop}%
\bibitem [{\citenamefont {Hauser}\ \emph {et~al.}(2016)\citenamefont {Hauser},
	\citenamefont {Richter}, \citenamefont {Homonnay}, \citenamefont
	{Eisenschmidt}, \citenamefont {Qaid}, \citenamefont {Deniz}, \citenamefont
	{Hesse}, \citenamefont {Sawicki}, \citenamefont {Ebbinghaus},\ and\
	\citenamefont {Schmidt}}]{Hauser2016}%
\BibitemOpen
\bibfield  {author} {\bibinfo {author} {\bibfnamefont {C.}~\bibnamefont
		{Hauser}}, \bibinfo {author} {\bibfnamefont {T.}~\bibnamefont {Richter}},
	\bibinfo {author} {\bibfnamefont {N.}~\bibnamefont {Homonnay}}, \bibinfo
	{author} {\bibfnamefont {C.}~\bibnamefont {Eisenschmidt}}, \bibinfo {author}
	{\bibfnamefont {M.}~\bibnamefont {Qaid}}, \bibinfo {author} {\bibfnamefont
		{H.}~\bibnamefont {Deniz}}, \bibinfo {author} {\bibfnamefont
		{D.}~\bibnamefont {Hesse}}, \bibinfo {author} {\bibfnamefont
		{M.}~\bibnamefont {Sawicki}}, \bibinfo {author} {\bibfnamefont {S.~G.}\
		\bibnamefont {Ebbinghaus}}, \ and\ \bibinfo {author} {\bibfnamefont
		{G.}~\bibnamefont {Schmidt}},\ }\href {\doibase 10.1038/srep20827} {\bibfield
	{journal} {\bibinfo  {journal} {Scientific Reports}\ }\textbf {\bibinfo
		{volume} {6}},\ \bibinfo {pages} {20827} (\bibinfo {year}
	{2016})}\BibitemShut {NoStop}%
\bibitem [{\citenamefont {Sun}\ \emph {et~al.}(2012)\citenamefont {Sun},
	\citenamefont {Song}, \citenamefont {Chang}, \citenamefont {Kabatek},
	\citenamefont {Jantz}, \citenamefont {Schneider}, \citenamefont {Wu},
	\citenamefont {Schultheiss},\ and\ \citenamefont {Hoffmann}}]{Sun2012}%
\BibitemOpen
\bibfield  {author} {\bibinfo {author} {\bibfnamefont {Y.}~\bibnamefont
		{Sun}}, \bibinfo {author} {\bibfnamefont {Y.-Y.}\ \bibnamefont {Song}},
	\bibinfo {author} {\bibfnamefont {H.}~\bibnamefont {Chang}}, \bibinfo
	{author} {\bibfnamefont {M.}~\bibnamefont {Kabatek}}, \bibinfo {author}
	{\bibfnamefont {M.}~\bibnamefont {Jantz}}, \bibinfo {author} {\bibfnamefont
		{W.}~\bibnamefont {Schneider}}, \bibinfo {author} {\bibfnamefont
		{M.}~\bibnamefont {Wu}}, \bibinfo {author} {\bibfnamefont {H.}~\bibnamefont
		{Schultheiss}}, \ and\ \bibinfo {author} {\bibfnamefont {A.}~\bibnamefont
		{Hoffmann}},\ }\href {\doibase 10.1063/1.4759039} {\bibfield  {journal}
	{\bibinfo  {journal} {Applied Physics Letters}\ }\textbf {\bibinfo {volume}
		{101}},\ \bibinfo {pages} {152405} (\bibinfo {year} {2012})}\BibitemShut
{NoStop}%
\bibitem [{\citenamefont {Pirro}\ \emph {et~al.}(2014)\citenamefont {Pirro},
	\citenamefont {Br{\"{a}}cher}, \citenamefont {Chumak}, \citenamefont
	{L{\"{a}}gel}, \citenamefont {Dubs}, \citenamefont {Surzhenko}, \citenamefont
	{G{\"{o}}rnert}, \citenamefont {Leven},\ and\ \citenamefont
	{Hillebrands}}]{Pirro2014}%
\BibitemOpen
\bibfield  {author} {\bibinfo {author} {\bibfnamefont {P.}~\bibnamefont
		{Pirro}}, \bibinfo {author} {\bibfnamefont {T.}~\bibnamefont
		{Br{\"{a}}cher}}, \bibinfo {author} {\bibfnamefont {A.~V.}\ \bibnamefont
		{Chumak}}, \bibinfo {author} {\bibfnamefont {B.}~\bibnamefont {L{\"{a}}gel}},
	\bibinfo {author} {\bibfnamefont {C.}~\bibnamefont {Dubs}}, \bibinfo {author}
	{\bibfnamefont {O.}~\bibnamefont {Surzhenko}}, \bibinfo {author}
	{\bibfnamefont {P.}~\bibnamefont {G{\"{o}}rnert}}, \bibinfo {author}
	{\bibfnamefont {B.}~\bibnamefont {Leven}}, \ and\ \bibinfo {author}
	{\bibfnamefont {B.}~\bibnamefont {Hillebrands}},\ }\href {\doibase
	10.1063/1.4861343} {\bibfield  {journal} {\bibinfo  {journal} {Applied
			Physics Letters}\ }\textbf {\bibinfo {volume} {104}},\ \bibinfo {pages}
	{012402} (\bibinfo {year} {2014})}\BibitemShut {NoStop}%
\bibitem [{\citenamefont {Hahn}\ \emph {et~al.}(2014)\citenamefont {Hahn},
	\citenamefont {Naletov}, \citenamefont {de~Loubens}, \citenamefont {Klein},
	\citenamefont {d'Allivy Kelly}, \citenamefont {Anane}, \citenamefont
	{Bernard}, \citenamefont {Jacquet}, \citenamefont {Bortolotti}, \citenamefont
	{Cros}, \citenamefont {Prieto},\ and\ \citenamefont
	{Mu{\~{n}}oz}}]{Hahn2014}%
\BibitemOpen
\bibfield  {author} {\bibinfo {author} {\bibfnamefont {C.}~\bibnamefont
		{Hahn}}, \bibinfo {author} {\bibfnamefont {V.~V.}\ \bibnamefont {Naletov}},
	\bibinfo {author} {\bibfnamefont {G.}~\bibnamefont {de~Loubens}}, \bibinfo
	{author} {\bibfnamefont {O.}~\bibnamefont {Klein}}, \bibinfo {author}
	{\bibfnamefont {O.}~\bibnamefont {d'Allivy Kelly}}, \bibinfo {author}
	{\bibfnamefont {A.}~\bibnamefont {Anane}}, \bibinfo {author} {\bibfnamefont
		{R.}~\bibnamefont {Bernard}}, \bibinfo {author} {\bibfnamefont
		{E.}~\bibnamefont {Jacquet}}, \bibinfo {author} {\bibfnamefont
		{P.}~\bibnamefont {Bortolotti}}, \bibinfo {author} {\bibfnamefont
		{V.}~\bibnamefont {Cros}}, \bibinfo {author} {\bibfnamefont {J.~L.}\
		\bibnamefont {Prieto}}, \ and\ \bibinfo {author} {\bibfnamefont
		{M.}~\bibnamefont {Mu{\~{n}}oz}},\ }\href {\doibase 10.1063/1.4871516}
{\bibfield  {journal} {\bibinfo  {journal} {Applied Physics Letters}\
	}\textbf {\bibinfo {volume} {104}},\ \bibinfo {pages} {152410} (\bibinfo
	{year} {2014})}\BibitemShut {NoStop}%
\bibitem [{\citenamefont {Dubs}\ \emph {et~al.}(2016)\citenamefont {Dubs},
	\citenamefont {Surzhenko}, \citenamefont {Linke}, \citenamefont {Danilewsky},
	\citenamefont {Br{\"{u}}ckner},\ and\ \citenamefont {Dellith}}]{Dubs2016}%
\BibitemOpen
\bibfield  {author} {\bibinfo {author} {\bibfnamefont {C.}~\bibnamefont
		{Dubs}}, \bibinfo {author} {\bibfnamefont {O.}~\bibnamefont {Surzhenko}},
	\bibinfo {author} {\bibfnamefont {R.}~\bibnamefont {Linke}}, \bibinfo
	{author} {\bibfnamefont {A.}~\bibnamefont {Danilewsky}}, \bibinfo {author}
	{\bibfnamefont {U.}~\bibnamefont {Br{\"{u}}ckner}}, \ and\ \bibinfo {author}
	{\bibfnamefont {J.}~\bibnamefont {Dellith}},\ }\href
{http://arxiv.org/abs/1608.08043} {(\bibinfo {year}
	{2016})},\ \Eprint {http://arxiv.org/abs/1608.08043} {arXiv:1608.08043}
\BibitemShut {NoStop}%
\bibitem [{\citenamefont {R{\"{o}}schmann}(1981)}]{Roschmann1981}%
\BibitemOpen
\bibfield  {author} {\bibinfo {author} {\bibfnamefont {P.}~\bibnamefont
		{R{\"{o}}schmann}},\ }\href {\doibase 10.1109/TMAG.1981.1061632} {\bibfield
	{journal} {\bibinfo  {journal} {IEEE Transactions on Magnetics}\ }\textbf
	{\bibinfo {volume} {17}},\ \bibinfo {pages} {2973} (\bibinfo {year}
	{1981})}\BibitemShut {NoStop}%
\end{thebibliography}
\end{document}